\documentclass[twocolumn,10pt,amsmath,amssymb,superscriptaddress,pra,aps]{revtex4-1}
\usepackage{graphicx}
\usepackage{hyperref}
\usepackage{charter}
\usepackage{amsmath}
\usepackage{amsfonts}
\usepackage{amssymb}%
\hypersetup{colorlinks,breaklinks,urlcolor=[rgb]{0.,0.4,0.4},linkcolor=blue,citecolor=red}
\begin{document}
\title{Capturing non-exponential dynamics in the presence of two decay channels}
\author{Francesco Giacosa}
\affiliation{Institute of Physics, Jan-Kochanowski University, ul. Uniwersytecka 7,
PL-25406, Kielce, Poland}
\affiliation{\mbox{Institute for Theoretical Physics, J. W. Goethe University, Max-von-Laue-Str. 1, DE-60438 Frankfurt am Main, Germany}}
\author{Przemys\l aw Ko\'{s}cik}
\affiliation{Department of Computer Science, State Higher Vocational School in
Tarn\'{o}w, ul. Mickiewicza 8, PL-33100 Tarn\'{o}w, Poland}
\author{Tomasz Sowi\'{n}ski}
\affiliation{Institute of Physics, Polish Academy of Sciences, Aleja Lotnik\'{o}w
32/46, PL-02668 Warsaw, Poland}

\begin{abstract}
The most unstable quantum states and elementary particles possess more than a
single decay channel. At the same time, it is well known that typically the
decay law is not simply exponential. Therefore, it is natural to ask how to
spot the non-exponential decay when (at least) two decay channels are opened.
In this work, we study the tunneling phenomenon of an initially localized
particle in two spatially opposite directions through two different barriers,
mimicking two decay channels. In this framework, through a specific quantum
mechanical examples which can be accurately solved, we study general
properties of a two-channel decay that apply for various unstable quantum
states (among which also for unstable particles). Apart from small deviations
at early times, the survival probability and the partial tunneling probability
along the chosen direction are very well described by the exponential-decay
model. In contrast, the ratios of the decay probabilities and probability currents are evidently not a simple constant (as they would be in the
exponential limit) but display time-persisting oscillations. Hence, these ratios
are optimal witnesses of deviations from the exponential decay law.
\end{abstract}
\maketitle
\affiliation{Institute of Physics, Jan-Kochanowski University, ul. Uniwersytecka 7,
PL-25406, Kielce, Poland}
\affiliation{Institute for Theoretical Physics, J. W. Goethe University, Max-von-Laue-Str.
1, DE-60438 Frankfurt am Main, Germany}
\affiliation{Department of Computer Science, State Higher Vocational School in Tarn\'{o}w,
ul. Mickiewicza 8, PL-33100 Tarn\'{o}w, Poland}
\affiliation{Institute of Physics, Polish Academy of Sciences, Aleja Lotnik\'{o}w 32/46,
PL-02668 Warsaw, Poland}

\section{Introduction}

The fact that the decay law in Quantum Mechanics (QM) is not described by an
exponential function is well-established
\cite{khalfin,fonda,winter,levitan,dicus,peshkin,wyrzykowski,koide,calderon,duecan,giacosapra1,raczynska,fpnew}%
. In particular, decaying systems very often exhibit the so-called
\textit{Zeno period} at short initial times, in which the nondecay probability, \textit{i.e.}, the probability $p(t)$ that the unstable particle prepared at the initial time $t=0$ has not decayed
yet at a later time $t>0$, is quadratic in time, $p(t)-1\propto -t^{2}$. On
the other hand, for very long times (typically several orders of magnitude larger than the lifetime \cite{fonda}), the nondecay probability is typically governed by a power law. From the experimental point of view, the
deviations from the exponential decay have been verified at short times in the
study of tunneling of sodium atoms in an optical potential \cite{raizen} and
more recently in the study of decays of unstable molecules via emission of
photons \cite{rothe}. Even if ubiquitous from a theoretical point of view, in
physical systems the deviations from the exponential case are typically very
small, making them very difficult to be measured.

Quite remarkably, the non-exponential decay allows also influencing the decay
rate by changing the way how the measurement is performed. As examples, the
famous Quantum Zeno effect (QZE) and the Inverse Zeno Effect (IZE) are direct
consequences of the peculiarity of the decay law
\cite{dega,misra,kk1,kk2,koshino,solokovski,fptopicalreview,fp,nakazato,facchiprl,schulman,giacosapra2,gpn}%
. Indeed, experimental confirmation of both the QZE and the IZE was
achieved in experiments in which electrons undergo Rabi transition between
atomic energy levels \cite{itano,balzer,streed}. In these cases, the nondecay
probability oscillates in time as $\sim\cos^{2}(\Omega t)$ and is evidently
non-exponential. Even if this is not a real unstable system, the slow-down of
the quantum transition by frequent measurements could be seen in these
experiments. Even more interestingly, these effects were also confirmed in the
tunneling of sodium atoms, which represent a genuine irreversible quantum
decay \cite{raizen2}. Finally, the QZE and IZE are also related to the quantum
computation and quantum control, which are important elements in this
flourishing research field \cite{lidar,qzs}.

Deviations from the exponential decay law are indeed expected also in Quantum
Field Theory (QFT), which is the ultimate correct framework to study the
creation and annihilation of particles, and hence the decay of unstable
particles \cite{duecan,vanhove,nonexpqft}. Namely, even if a perturbative
treatment is not capable to capture such deviations \cite{maiani}, the
spectral function in QFT is not a Breit-Wigner \cite{salam,salam2,lupo} and,
in some cases, it can be very much different from it \cite{pelaezrev}. Then,
as a consequence, also the decay law is not a simple exponential. Unfortunately, a direct experimental proof of the nonexponential decay of unstable elementary particles is still missing. Nonetheless, the Zeno effect confirmed recently in cavity QED \cite{haroche} suggests that different dynamical features of the simplest QM systems may have their counterparts also in different purely QFT situations.

An interesting case is realized when an unstable quantum state (or particle)
can decay in (at least) two channels. Indeed, this situation takes place very
often in Nature. For instance, in the realm of particle physics, most unstable
particles posses multiple decay channels \cite{pdg}. Similarly, electrons in
excited atoms can decay in more than into a single energy level \cite{scully}.

As expected, in the exponential limit, the ratio of the decay probabilities into the first and the second channel is a constant. A detailed study of the non-exponential decay when two (or more) decay channels are present is described in \cite{duecan}. In QM, this ratio is not a constant but
shows some peculiar and irregular oscillations, which in \cite{duecan} were
discussed in the framework of the so-called Lee model \cite{lee,lee2} (also
called the Friedrichs model or the Jaynes-Cummings model \cite{scully,sherman}%
) which captures the most salient features of QFT (for details see
\cite{duecan,qcdeff,xiaozhou,xiaozhou2,pok}). Moreover, a qualitatively
similar results for the ratio of the partial decay probability
currents were obtained in \cite{duecan} also in a quantum field
theoretical model. Yet, the topic of non-exponential decay in the presence of
more decay channel needs novel and different studies that allow us to
understand more in detail its features and to make an experimental
verification (or falsification) possible.

In this work, we intend to explore the two-channel decay problem in a quantum mechanical context. To this aim, we introduce a simple model of a single-particle initially confined in a box potential whose walls are suddenly partially released allowing the particle to tunnel to the open space. In this way we slightly generalize the celebrated \textit{Winter's model} \cite{winter} where only a single box wall is released. The Winters model is recognized as one of the most important workhorses in the theory of non-exponential decays (see for example \cite{levitan,dicus,peshkin,wyrzykowski,koide,calderon} and \cite{razavy} for a general treatment). In our work we want to mimic two different channels of a decay and therefore we focus on situations of essentially different barriers. In contrast to the symmetric situation of identical barriers \cite{Bauch,Demkov,Kleber}, in this case the exact analytical solution is known only for the scattering problem of external wave packets \cite{Guilarte,ahmed,nogamiross,yanetka,erman} and it does not provide straightforward solution for the decay scenario studied here \cite{footnote2}. Solving for all practical purposes the corresponding time-dependent Schr\"{o}dinger equation exactly (in numerical means) we check how to capture deviations from the exponential decay law. In agreement with Ref.
\cite{duecan}, but with a different method, we find that the ratio of the decay probability currents shows time-persisting deviations from the
exponential decay law predictions. The main advantage of the approach
presented here is its complete transparency of all successive steps and its
feasibility in physical experiments in which the tunneling in different directions can be obtained by asymmetric potentials. Moreover, as discussed in the summary, the qualitative features of the obtained results are expected to be quite general and can be used not only to describe generic tunneling processes of particles to the open space but also to understand decays of unstable relativistic particles in the QFT language.

\section{The model}

In this paper we consider a single particle moving in a one-dimensional space
subjected to two separated delta potential barriers. The system is described
by the following Hamiltonian
\begin{equation}
\mathcal{H}=-\frac{\hbar^{2}}{2m}\frac{\mathrm{d}^{2}}{\mathrm{d}x^{2}}%
+V_{L}\delta(x+R)+V_{R}\delta(x-R)\text{ ,} \label{hamilton}%
\end{equation}
where $R$ is the half-distance between the two barriers and their height is
controlled by the independent parameters $V_{L}$ and $V_{R}$. Our aim
is to find the decay properties of a particle that is initially located
between the barriers. To this aim, at the initial moment ($t=0$) the wave
function is taken as
\begin{equation}
\Psi(x,t=0)=\Psi_{0}(x)=\left\{
\begin{array}
[c]{ll}%
\frac{1}{\sqrt{R}}\cos\left( \frac{\pi x}{2R}\right) & |x|\leq R\\
0 & |x|>R
\end{array}
\right. \text{ ,} \label{initial}%
\end{equation}
which corresponds to the ground state in the limit of barriers of infinite
heights. This choice is quite natural, but of course one could use other
initial wave functions without changing the qualitative results that we are
going to present.

The properties of the system studied are controlled by only two independent
dimensionless parameters. It is clearly visible that all quantities can be
expressed in units fixed by the half-distance $R$. Namely, if all distances
are measured in unit of $R$, energies in unit of $\hbar^{2}/(mR^{2})$, and
time intervals in unit of $mR^{2}/\hbar$, then the properly rescaled
(dimensionless) Hamiltonian takes the form
\begin{equation}
\mathcal{H}=-\frac{1}{2}\frac{\mathrm{d}^{2}}{\mathrm{d}x^{2}}+V_{0}\left[
\delta(x+1)+\kappa\delta(x-1)\right] , \label{hamiltonadim}%
\end{equation}
where $V_{0}=\frac{mR}{\hbar^{2}}V_{L}$ and $\kappa=V_{R}/V_{L}$ are two
independent dimensionless parameters controlling the heights of the left
barrier and the ratio between the right and the left heights, respectively. In
these units, we solve the time-dependent Schr\"{o}dinger equation
\begin{equation}
\left( i\partial_{t}-\mathcal{H}\right) \Psi(x,t)=0 \label{se}%
\end{equation}
with the initial wave function (\ref{initial}). Notice that, in the chosen
units, the initial energy of the system (in the limit $V_{0}\rightarrow
\infty,$ and $\kappa>0$) is $E_{0}=\pi^{2}/8$, which is of order $1$. Clearly,
due to the mirror symmetry of the problem, without losing generality, one can
restrict to $0<\kappa\leq1$.

To quantify the dynamics of the system we focus our attention on \textit{the
nondecay probability} defined as
\begin{equation}
\mathrm{P}_{0}(t)=\int_{-1}^{+1}\!\mathrm{d}x\,|\Psi(x,t)|^{2}, \label{ndp}%
\end{equation}
\textit{i.e.}, the probability that the particle is remaining in the region
$x\in(-1,1)$ at the time $t$. Note that this quantity is interchangeably also
called \textit{the survival probability}, but then some attention is needed
\cite{footnote}. Moreover, we also consider the left and the right decay
probabilities defined as
\begin{subequations}
\label{Prob}%
\begin{align}
\mathrm{P}_{L}(t) & =\int_{-\infty}^{-1}\!\mathrm{d}x\,|\Psi(x,t)|^{2},\\
\mathrm{P}_{R}(t) & =\int_{+1}^{+\infty}\!\mathrm{d}x\,|\Psi(x,t)|^{2},
\end{align}
where $\mathrm{P}_{L}(t)$ ($\mathrm{P}_{R}(t)$) is the probability that at the
time $t$ the particle can be found to the left (right) of the well,
\textit{i.e.}, it is the probability that the tunneling to the left (right)
has occurred in the time interval between $0$ and $t$. Obviously, at any
instant $t$ these probabilities are not independent and must obey the
normalization condition
\end{subequations}
\begin{equation}
\mathrm{P}_{0}(t)+\mathrm{P}_{L}(t)+\mathrm{P}_{R}(t)=1\text{ .} \label{sum}%
\end{equation}
It is also extremely useful to consider the probability currents (the time derivatives of the probabilities) describing the speed of their temporal change:
\begin{equation}
\mathrm{p}_{0}(t)=-\frac{\mathrm{d}\mathrm{P}_{0}(t)}{\mathrm{d}t}\text{
,}\,\mathrm{p}_{L}(t)=\frac{\mathrm{d}\mathrm{P}_{L}(t)}{\mathrm{d}t}\text{
,}\,\mathrm{p}_{R}(t)=\frac{\mathrm{d}\mathrm{P}_{R}(t)}{\mathrm{d}t}.
\label{densprobdef}%
\end{equation}
Notice that the definition of $\mathrm{p}_{0}(t)$ takes into account that the
nondecay probability decreases with time. Temporal changes of $\mathrm{p}%
_{0}(t)$ are often measured in experiments, since it corresponds to the number
of decay products per unit time (for instance the lifetime measurement of the
neutron by the beam method \cite{neutronbeam} or the decay of H-like ions via
electron capture and neutrino emission \cite{gsilast}). Note, a simple
interpretation holds: $\mathrm{p}_{L(R)}(t)\mathrm{d}t$ is the probability
that the decay occurs to the right (left) between $t$ and $t+\mathrm{d}t.$
Clearly, from the relation (\ref{sum}) one finds that
\begin{equation}
\mathrm{p}_{0}(t)=\mathrm{p}_{L}(t)+\mathrm{p}_{R}(t).
\end{equation}
The central quantities we focus in the following are the right-to-left ratio
of probabilities
\begin{equation}
\boldsymbol{\Pi}(t)=\frac{{\mathrm{P}_{R}(t)}}{{\mathrm{P}_{L}(t)}}
\label{ratio1}%
\end{equation}
and its counterpart, the right-to-left ratio of probability currents
\begin{equation}
\boldsymbol{\pi}(t)=\frac{{\mathrm{p}_{R}(t)}}{{\mathrm{p}_{L}(t)}}\text{ .}
\label{ratio2}%
\end{equation}
It will turn out that time-dependence of both ratios plays a crucial role in
capturing non-exponential decay behavior of the system.

Finally, let us recall the explicit forms of all these functions when the
exponential Breit-Wigner limit (BW) \cite{ww,ww2,ww3} holds. In this limit the
nondecay probability reads
\begin{equation}
\mathrm{P}_{0}(t)\xrightarrow{\text{BW}}e^{-\Gamma t} \label{exp}%
\end{equation}
where $\Gamma$ is the decay rate. As argued in \cite{fonda}, the exponential
dependence of the nondecay probability is a direct consequence of the
Breit-Wigner energy distribution of the unstable state. The decay rate
$\Gamma$ can be also decomposed to partial decay rates to the `left'
$\Gamma_{L}$ and to the `right' $\Gamma_{R}$ associated with these two
distinguished decay channels, $\Gamma=\Gamma_{L}+\Gamma_{R}$. Then, the
partial decay probabilities have the form
\begin{subequations}
\label{approx}%
\begin{align}
& \mathrm{P}_{L}(t)\xrightarrow{\text{BW}}\frac{\Gamma_{L}}{\Gamma}\left(
1-e^{-\Gamma t}\right) \text{ ,}\label{pllim}\\
& \mathrm{P}_{R}(t)\xrightarrow{\text{BW}}\frac{\Gamma_{R}}{\Gamma}\left(
1-e^{-\Gamma t}\right) . \label{prrim}%
\end{align}
Obviously, the partial decay probability currents read
\end{subequations}
\begin{equation}
{\mathrm{p}_{L}(t)}\xrightarrow{\text{BW}}\text{ }\Gamma_{L}e^{-\Gamma
t},\,\,{\mathrm{p}_{R}(t)}\xrightarrow{\text{BW}}\text{ }\Gamma_{R}e^{-\Gamma
t}.
\end{equation}
For future convenience, we introduce the ratio of the partial decay widths
\begin{equation}
\beta=\Gamma_{R}/\Gamma_{L}
\end{equation}
which in the BW limit remains constant and it directly connects the right-to-left ratios (\ref{ratio1}) and (\ref{ratio2})
\begin{equation}
\boldsymbol{\Pi}(t)=\frac{{\mathrm{P}_{R}(t)}}{{\mathrm{P}_{L}(t)}%
}\xrightarrow{\text{BW}}\beta\xleftarrow{\text{BW}}\frac{{\mathrm{p}_{R}(t)}%
}{{\mathrm{p}_{L}(t)}}=\boldsymbol{\pi}(t)\text{ .} \label{explimitratios}%
\end{equation}
To show that the exponential decay law is violated it is sufficient to expose deviations from the constant value of $\beta=\Gamma_{R}/\Gamma_{L}$. This is why the right-to-left ratios (\ref{ratio1}) and (\ref{ratio2}) are of special interest.

\section{ Results}

We solve the Schr\"{o}dinger equation (\ref{se}) by expressing
the time-dependent wave function in terms of eigenstates of the dimensionless
Hamiltonian (\ref{hamiltonadim}). In practice, due to a lack of convenient exact analytical solutions, we diagonalize it on a finite
spatial interval with closed boundaries at $x=\pm L$ with $L/R \gg1$ (for more
technical details see the Appendix). We then calculate the nondecay
probability $P_{0}(t)$, the partial decay probabilities $\mathrm{P}_{L}(t)$
and $\mathrm{P}_{R}(t)$, and finally the two ratios $\boldsymbol{\Pi}(t)$ and
$\boldsymbol{\pi}(t)$.

\begin{figure}[ptb]
\includegraphics[width=0.241\textwidth]{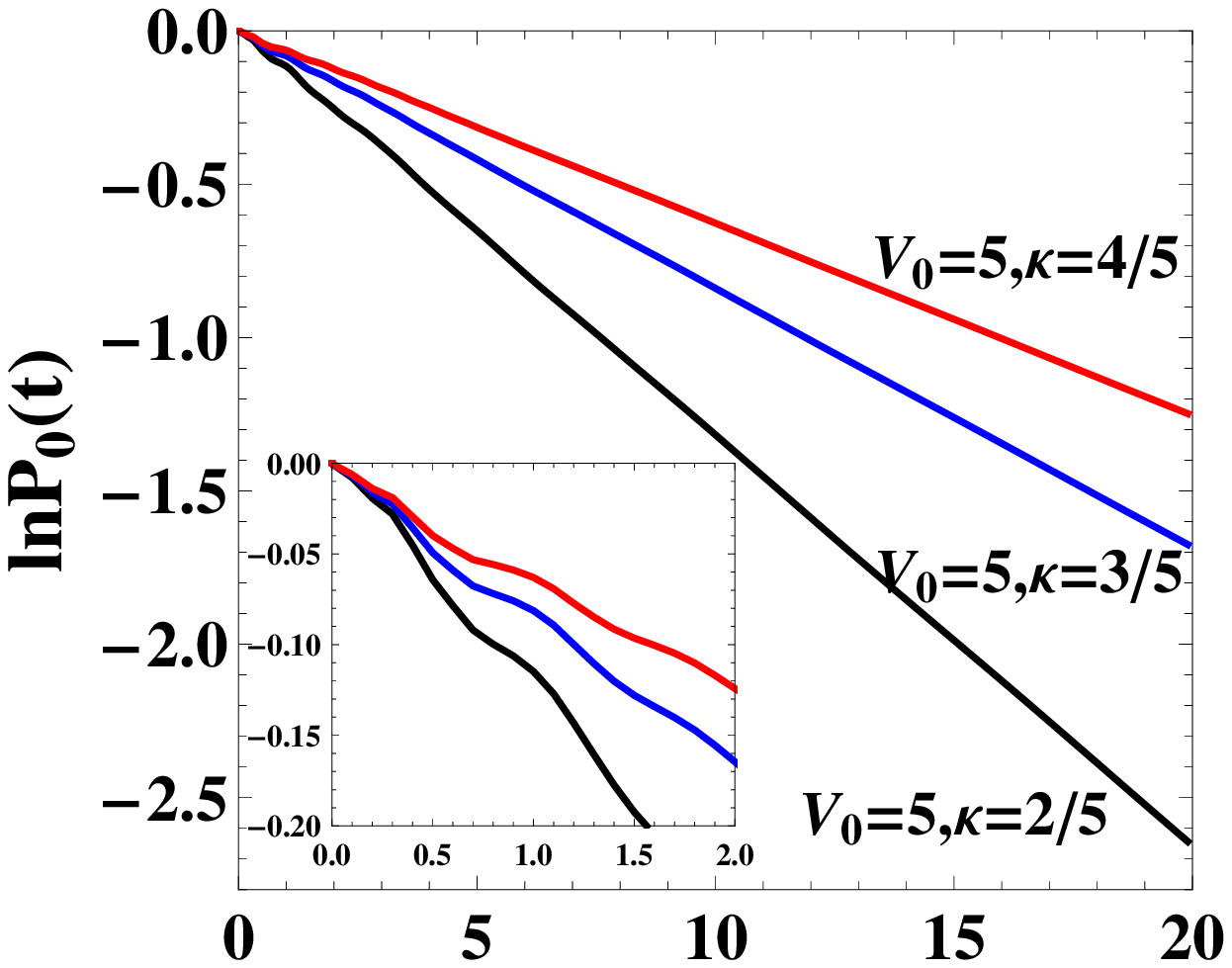}
\includegraphics[width=0.2365\textwidth]{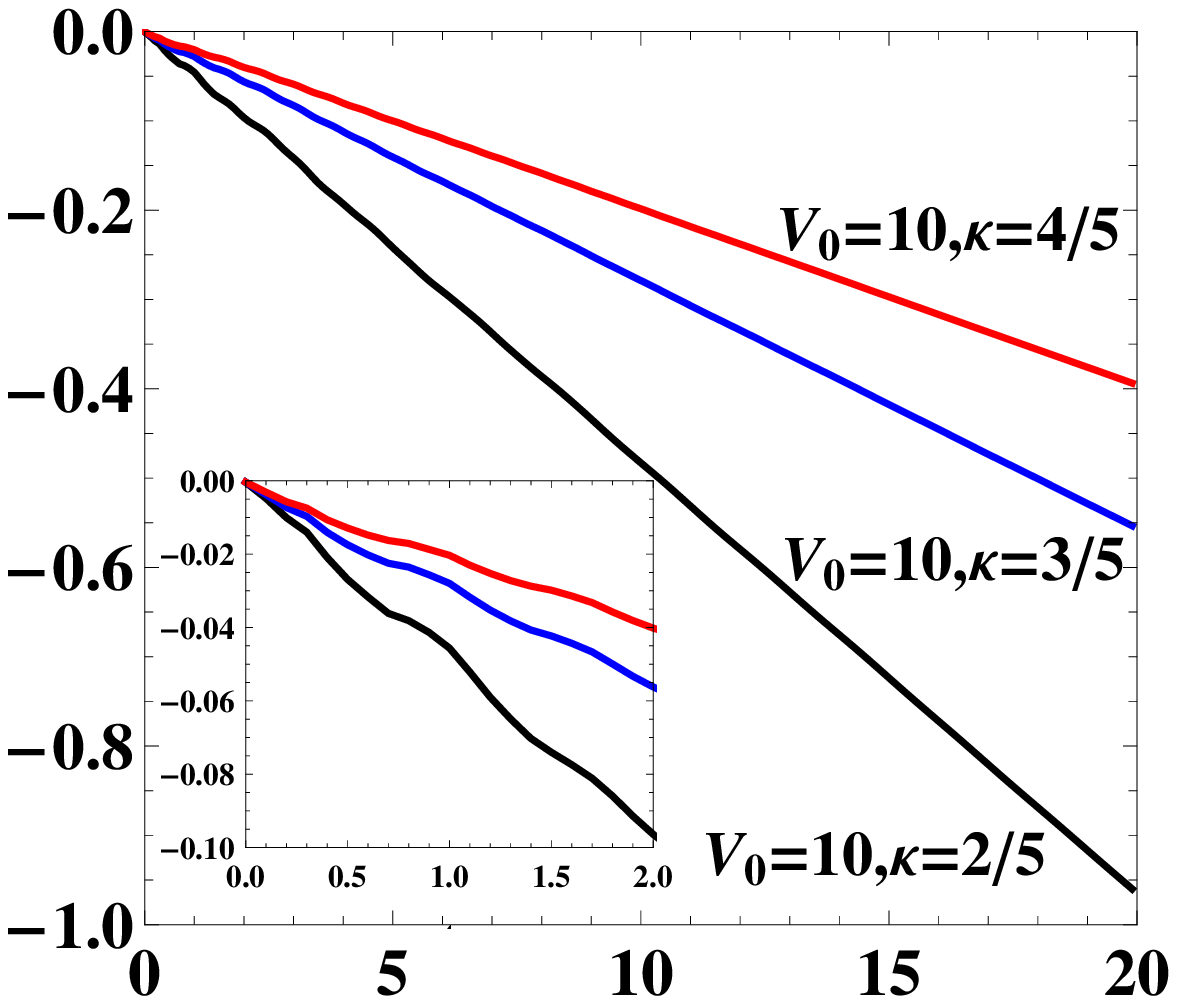}
\includegraphics[width=0.241\textwidth]{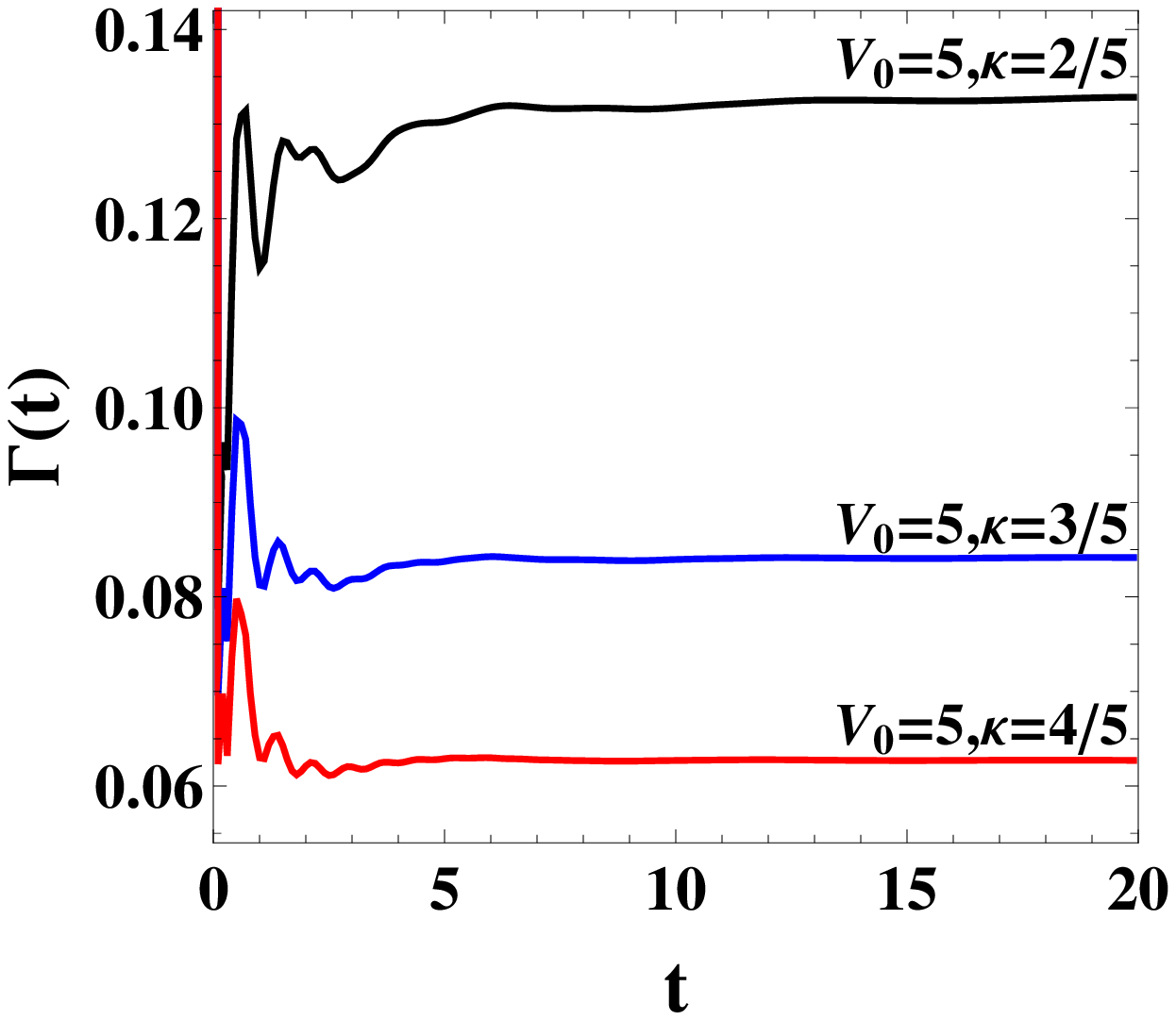}
\includegraphics[width=0.2365\textwidth]{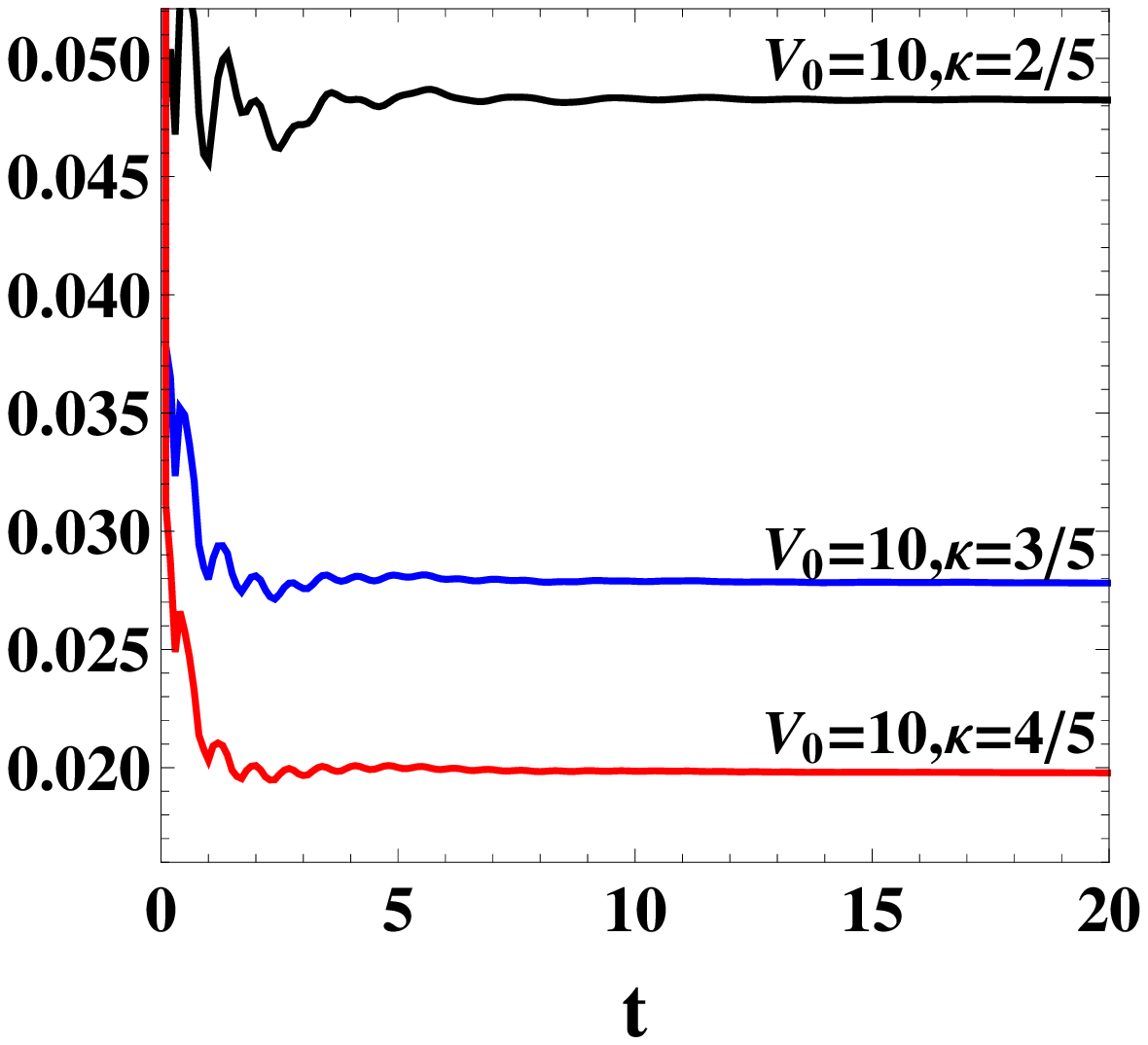}\caption{\textbf{Upper
panel}: the nondecay probability $\mathrm{P}_{0}(t)$ as a function of time for
some chosen values of $\kappa$ and $V_{0}$. The insets highlight the behavior
at short times. \textbf{Bottom panel}: The corresponding results for the decay
rate $\Gamma(t)=-{\mathrm{ln}\mathrm{P}_{0}(t)/t}$.}%
\label{Fig1}%
\end{figure}

\begin{figure}[ptb]
\includegraphics[width=0.485\textwidth]{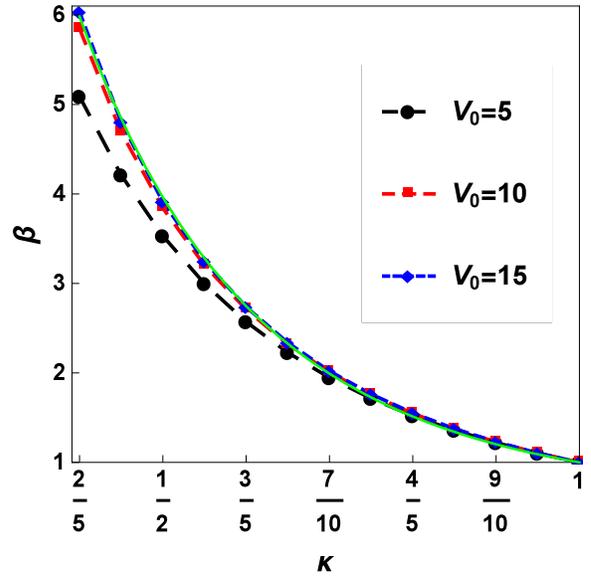}\caption{The ratio of
partial decay rates $\beta$ calculated in the BW limit as function of the
asymmetry parameter $\kappa$ for different values of $V_{0}$. The green solid
line indicates a phenomenological relation $\beta=\kappa^{-2}$ justified in
the limit of large $V_{0}$. }%
\label{Fig2}%
\end{figure}

In the upper panel of Fig.~\ref{Fig1} we show the nondecay probability
$\mathrm{P}_{0}(t)$ as a function of time for some chosen values of $V_{0}$
and $\kappa$ (the insets highlight the changes for small $t$). It is clearly
seen that, after a short initial period, $\mathrm{P}_{0}(t)$ exhibits an
exponential decay. It is even more evident when the decay rate $\Gamma
(t)=-{\mathrm{ln}\mathrm{P}_{0}(t)/t}$ is plotted (bottom panel in
Fig.~\ref{Fig1}) -- after some small initial wiggles, it reaches a constant
value indicating a quite fast transition to the BW regime. These results
suggest that in the regime of exponential decay the approximation (\ref{exp})
should be applied. It turns out that in this regime, the nondecay probability
almost ideally fits the relation
\begin{equation}
\mathrm{P}_{0}(t)\approx e^{-\Gamma(t-t_{0})} \label{expshifted}%
\end{equation}
manifesting the correctness of the BW limit predictions. Note, that in general
the additional \textquotedblleft time-shift\textquotedblright\ $t_{0}$ is
non-zero and its inverse is directly related to the initial period of
non-exponential decay. In fact, the sign of $t_{0}$ indicates if for small
times the dynamics is sub- or sup-exponential (see \cite{fptopicalreview} and \cite{sherman} for detailed discussions of this point). In the cases studied here, this parameter
is very close to $0$ and, due to numerical uncertainty, we are not able to
determine its sign. To gain a deeper insight into the validity of the BW
approximation, we additionally check how the ratio of partial decay rates
$\beta$ depends on $\kappa$ and $V_{0}$ (see Fig.~\ref{Fig2}). It turns out
that the ratio $\beta$ becomes insensitive to changes in $V_{0}$ when $V_{0}$
is large enough. In fact, for a considered range of $\kappa$, the changes in
$V_{0}$ do not affect the value of $\beta$ when $V_{0}$ exceeds a value of
about $15$. Moreover, in this regime the ratio $\beta$, when treated as a
function of $\kappa$, almost perfectly follow the simple relation
$\beta(\kappa)\approx\kappa^{-2}$ (green line in Fig.~\ref{Fig2}). This
relation has a direct intuitive phenomenological explanation. For large
$V_{0}$ tunnelings in opposite directions become almost independent and
therefore the ratio of tunneling amplitudes is simply given by the ratio of
the barrier heights, $\kappa^{-1}$. It means, that the ratio of probabilities
is controlled solely by $\kappa^{-2}$.

The discussion above means that the exponential formula provides a very good
approximation for large enough (but not too large) times. Clear
deviations are visible only for initial moments (for the cases studied
$t\lesssim5$). Of course, the deviations become larger for smaller $V_{0}$.
However, we focus on the cases in which $P_{0}(t)$ is almost exponential,
since this is the typically realized scenario in Nature. 

\begin{figure}[ptb]
\includegraphics[width=0.241\textwidth]{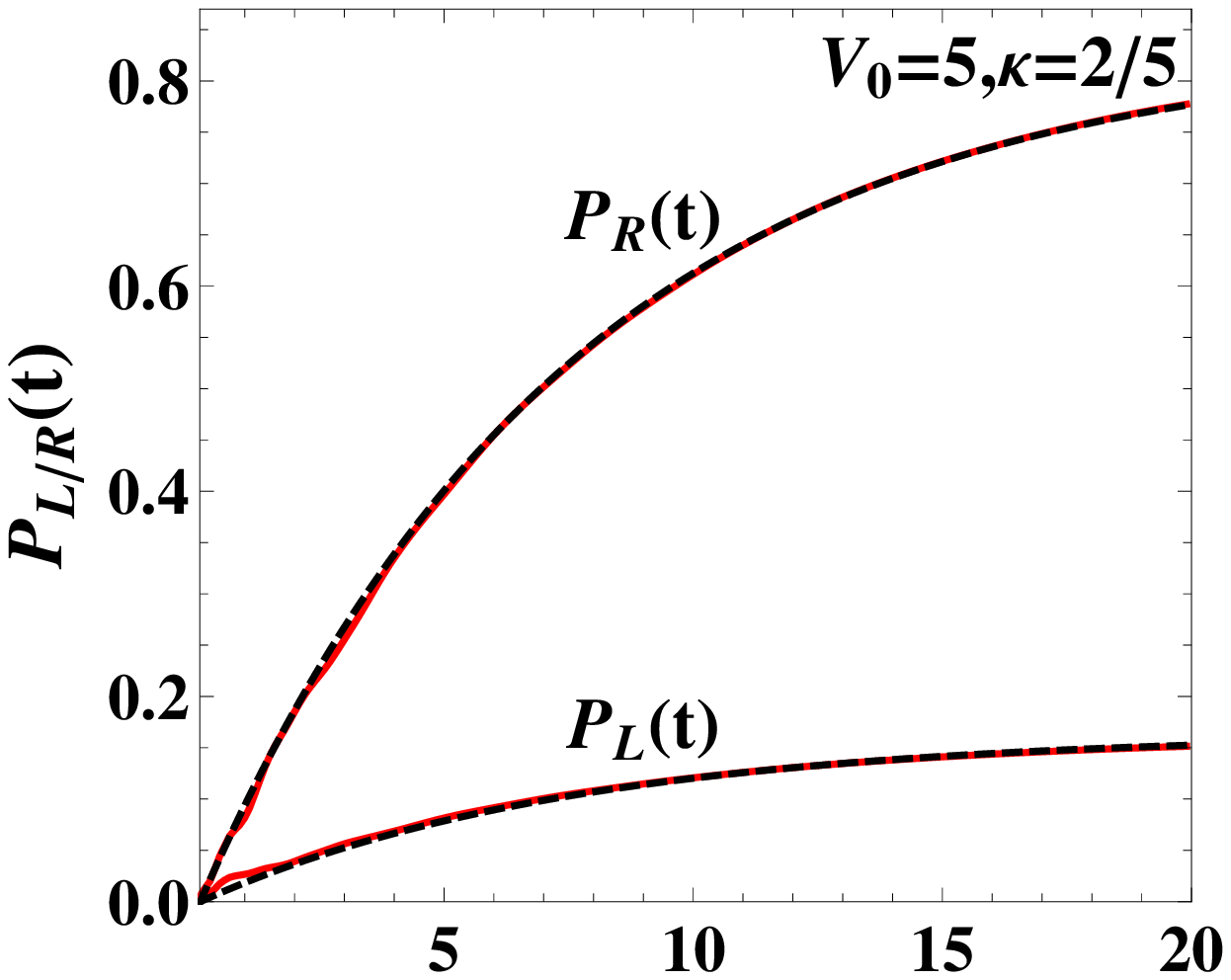}
\includegraphics[width=0.2365\textwidth]{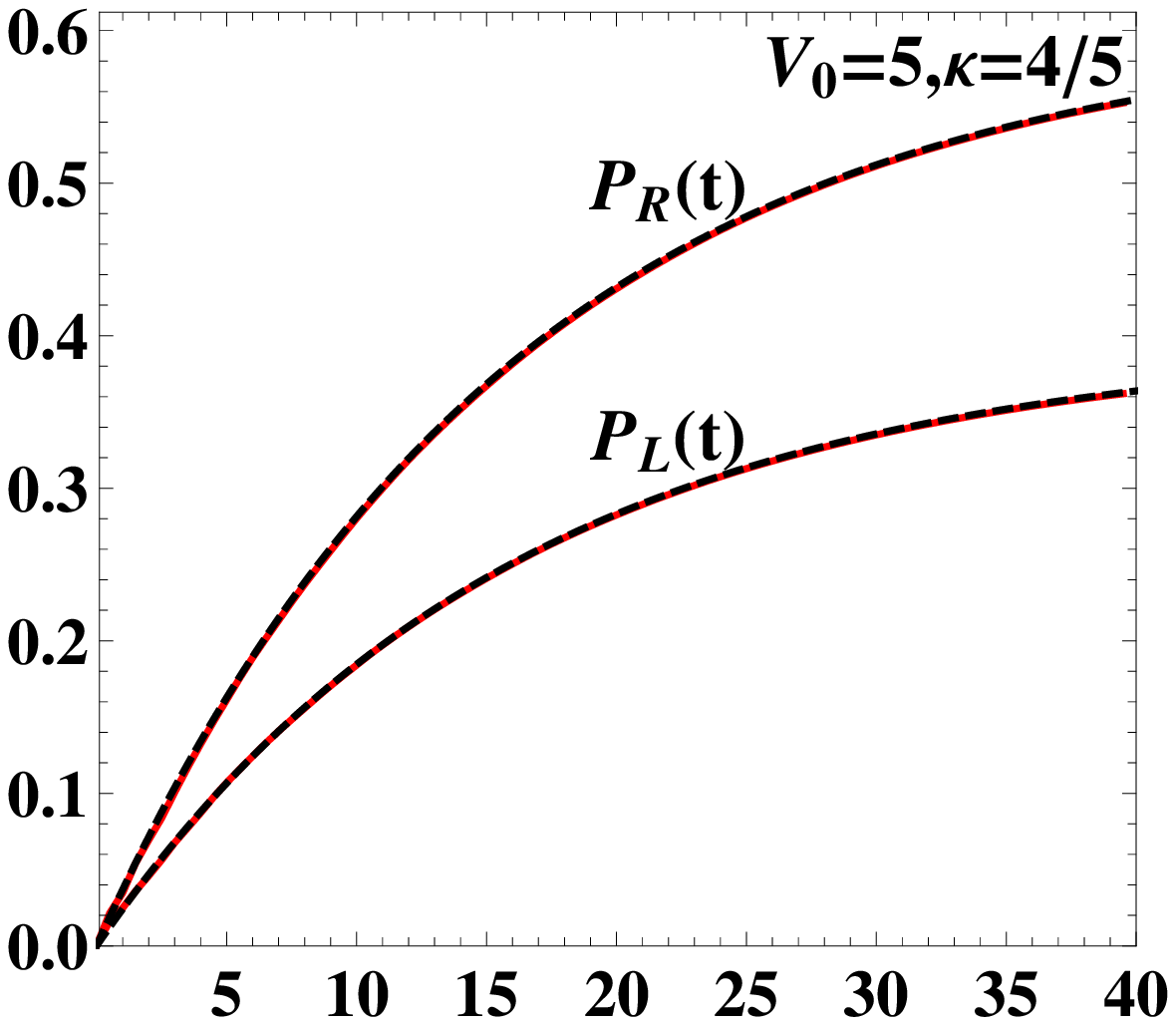}
\includegraphics[width=0.241\textwidth]{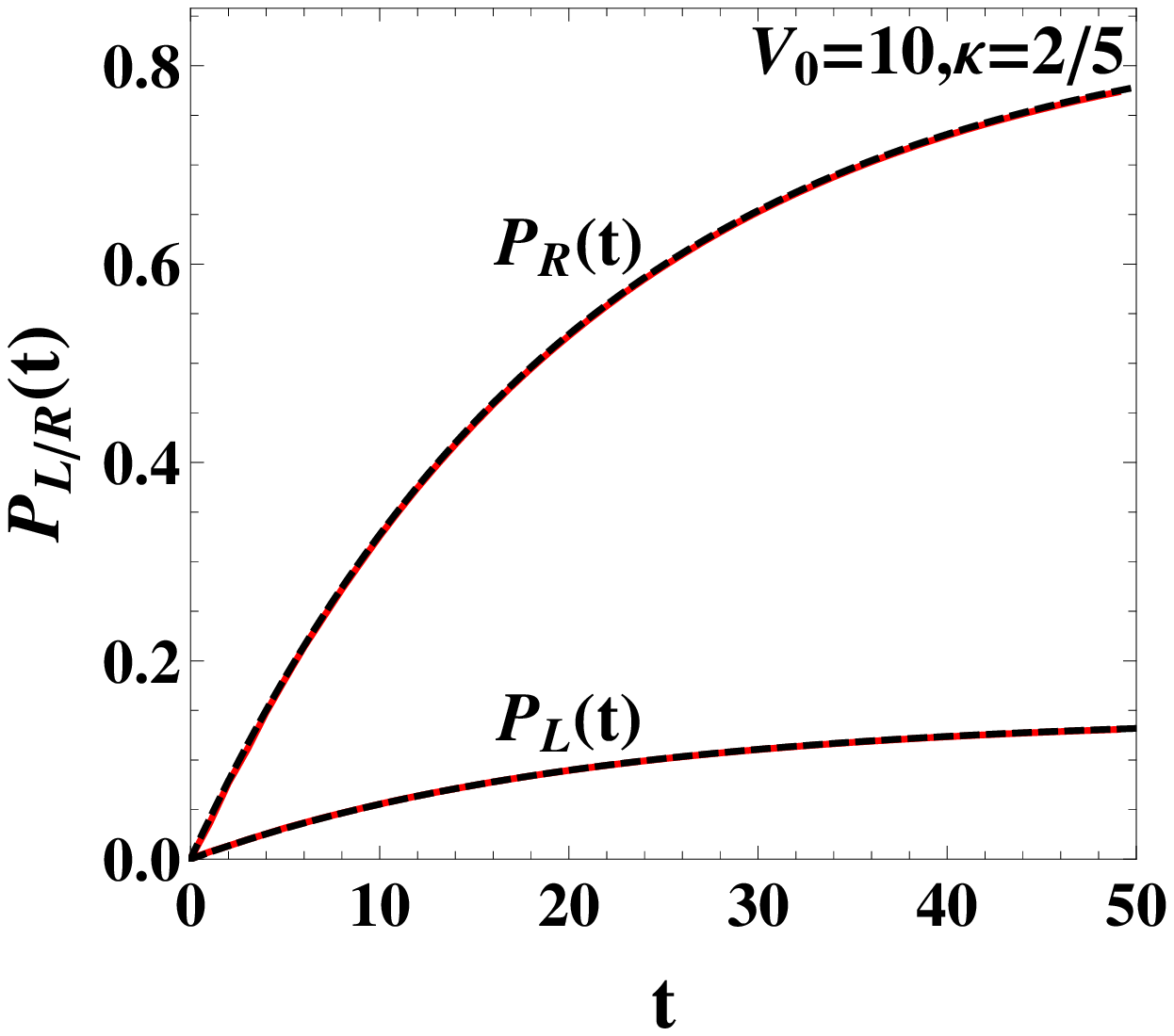}
\includegraphics[width=0.2365\textwidth]{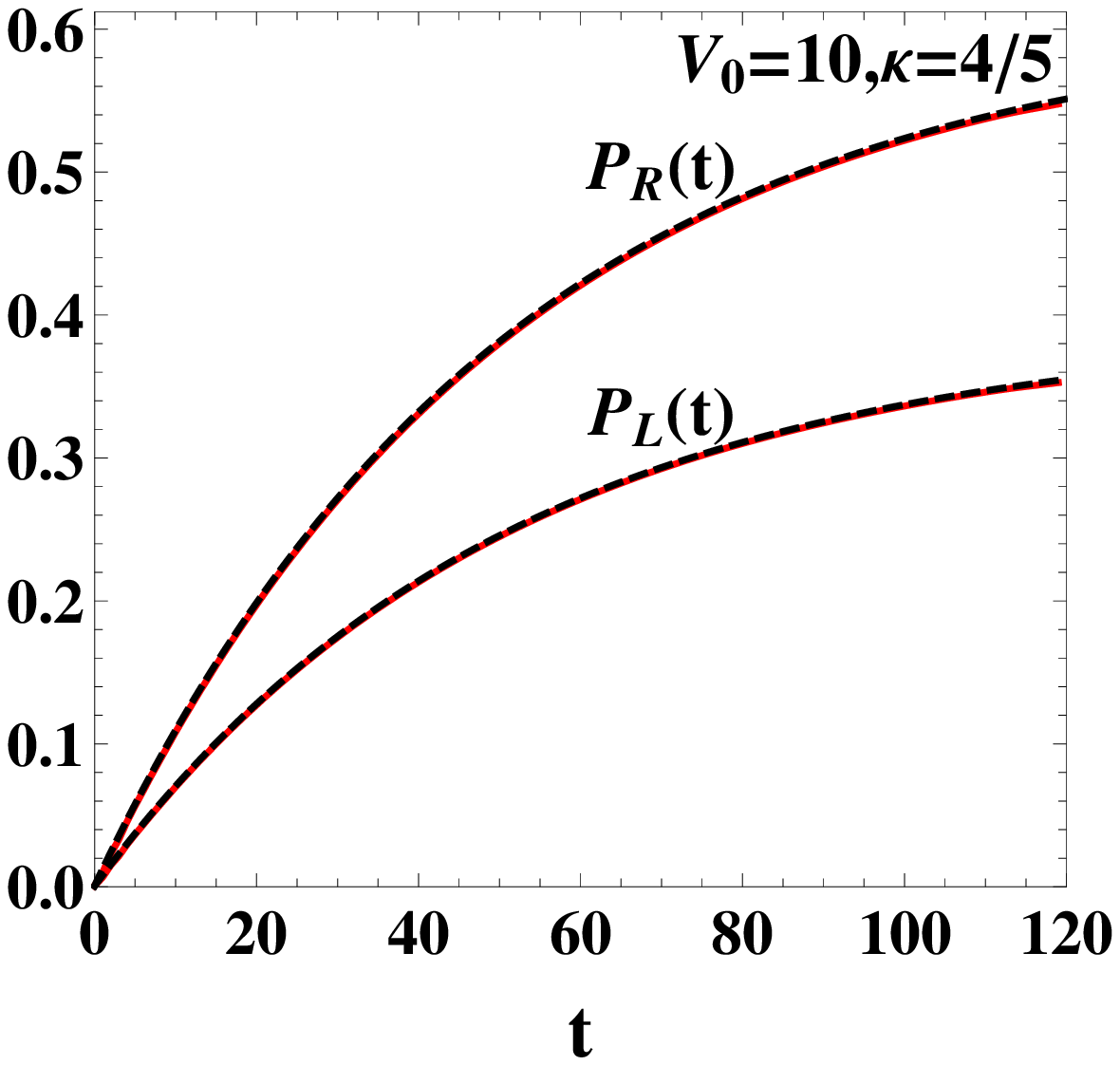}\caption{Partial decay
probabilities ${\mathrm{P}_{R}(t)}$ and ${\mathrm{P}_{L}(t)}$ as function of
time. Note that, accurate numerical results (continuous black lines)
coincidence with predictions of the BW limit (\ref{approx}) (red dashed
lines). See the main text for details.}%
\label{Fig3}%
\end{figure}

The situation is very similar when partial decay probabilities (\ref{Prob})
are considered. In this case, after fitting to appropriate exponential
functions of the form
\begin{equation}
\mathrm{P}_{L/R}(t)\approx\frac{\Gamma_{L/R}}{\Gamma}\left[ 1-e^{-\Gamma
(t-t_{0})}\right] \text{ ,} \label{expshifted2}%
\end{equation}
we see full agreement of the BW limit predictions with accurate numerical
results (see Fig.~\ref{Fig3} for comparison). 

\begin{figure}[ptb]
\includegraphics[width=0.2412\textwidth]{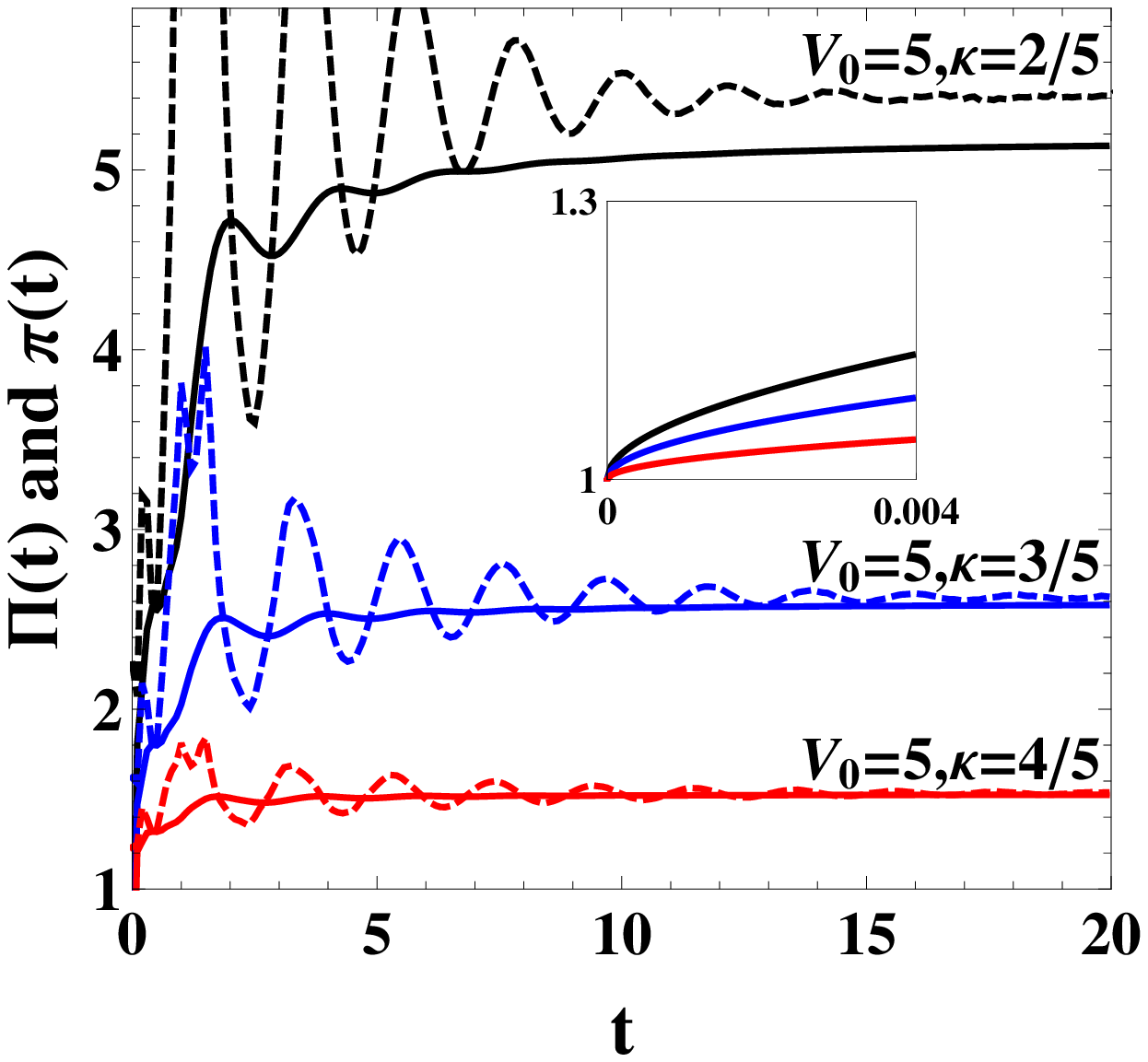}
\includegraphics[width=0.2363\textwidth]{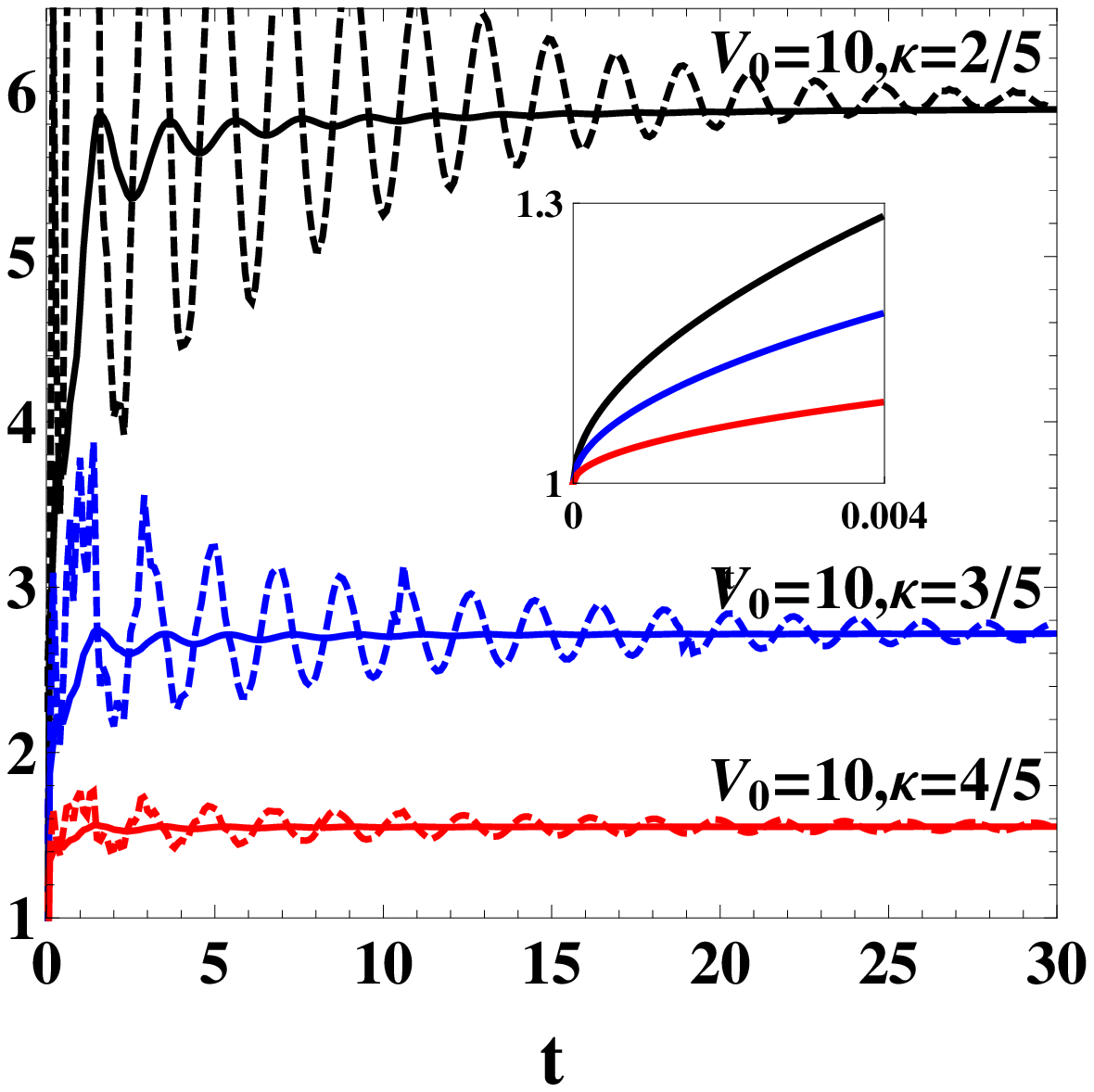}\caption{Temporal ratio of
partial probabilities $\boldsymbol{\Pi}(t)$ (continuous lines) and partial probability currents $\boldsymbol{\pi}(t)$ (dashed lines) as functions of time for the same set of parameters as in Fig.~\ref{Fig1}. The insets highlight the short-time behavior. Both quantities oscillate at intermediate times but the ratio $\boldsymbol{\pi}(t)$ shows evident deviations from the BW limit predictions even for very long times. }%
\label{Fig4}%
\end{figure}

All three results presented for probabilities $\mathrm{P}_{0}(t)$,
$\mathrm{P}_{R}(t)$, and $\mathrm{P}_{L}(t)$ suggest that any discrepancies
from the exponential behavior are poorly captured by these quantities. We
checked, that also this is the case when the probability currents
(\ref{densprobdef}), \textit{i.e.}, the temporal derivatives of the
probabilities, are considered. However, the situation changes dramatically
when, instead of pure probabilities (probability currents), the properties of
their temporal ratios $\boldsymbol{\Pi}(t)$ and $\boldsymbol{\pi}(t)$ are
investigated. In Fig.~\ref{Fig4} we present accurate numerical results for
these ratios as function of time for the same set of parameters as in
Fig.~\ref{Fig1}. One can see that the ratios $\boldsymbol{\Pi}(t)$ and
$\boldsymbol{\pi}(t)$ have rather complex behavior, especially for the initial
period. More importantly, the deviations from the constant value obtained in
the exponential BW limit are clearly visible. Both functions eventually reach
the expected constant value of $\beta$ in the limit of large times. Note however, that here we
do not consider very large times in which the decay is again non-exponential due to the onset of a power-law. In our studies, when referring to intermediate and large times, we mean periods in which the decay is almost ideally exponential.

In fact, our results allow us conclude that partial probabilities
$\mathrm{P}_{L}(t)$ and $\mathrm{P}_{R}(t)$ are generally linearly independent
functions since if $\boldsymbol{\Pi}(t)$ and $\boldsymbol{\pi}(t)$ are not
identically equal, then the Wronskian ${W}(t)=\mathrm{P}_{L}(t)\mathrm{p}%
_{R}(t)-\mathrm{P}_{R}(t)\mathrm{p}_{L}(t)$ is not singular. (Note, for
$\kappa=1$ symmetric tunneling to the left and to the right occurs:
$\boldsymbol{\Pi}(t)=\boldsymbol{\pi}(t)=1$). Only for a very large time, when
both ratios reach almost constant value $\beta$, one finds that
$\boldsymbol{\Pi}(t)-\boldsymbol{\pi}(t)\approx0$ which means that partial
probabilities $\mathrm{P}_{R}(t)$ and ${\mathrm{P}_{L}(t)}$ behave nearly as
linear dependent functions.

In particular, the right-to-left probability currents ratio $\boldsymbol{\pi
}(t)$ shows evident oscillations persisting for a very long time. It means
that it is an appropriate quantity to exhibit deviations from the
exponential BW limit predictions even in moments when the standard nondecay
probability ${\mathrm{P}_{0}(t)}$, the partial decay probabilities
${\mathrm{P}_{R}(t)}$ and ${\mathrm{P}_{L}(t)}$, or even their ratio
$\boldsymbol{\Pi}(t)$ are not able to capture this behavior. Let us also
recall that the ratio $\boldsymbol{\pi}(t)$ has a straightforward physical
meaning. For the time intervals in which $\boldsymbol{\pi}(t)>\beta$
($\boldsymbol{\pi}(t)<\beta$) the particle decay to the right is more (less)
probable than naively expected from the exponential law. Then, the value of
$\beta$ has only an appropriate interpretation as an average ratio. Closer
inspection of Fig.~\ref{Fig4} shows additional interesting insights for the
function $\boldsymbol{\pi}(t)$. Namely, the amplitude of oscillations does not
decrease in the limit of large $V_{0}$ as long as $\kappa$ is sufficiently
different from unity. Namely, when it approaches $1$, the ratio
$\boldsymbol{\pi}(t)$ rapidly flattens around the expected value $1$.
Consequently, in these cases, the deviations from the expected constant limit
become very small.

The above analysis shows that the ratios $\boldsymbol{\Pi}(t)$ and $\boldsymbol{\pi}(t)$ can be regarded as appropriate quantities capturing non-exponential decay in the presence of two decay channels. However, as we argued the ratio of the time derivatives $\boldsymbol{\pi}(t)$ is much more sensitive to non-exponential features of the system than the direct ratio of probabilities $\boldsymbol{\Pi}(t)$. Therefore, from the experimental point of view, if one aims to validate exponential decay, the largest effort should be put on accurate determination of the quantity $\boldsymbol{\pi}(t)$ rather then $\boldsymbol{\Pi}(t)$. 

It is interesting to note that for a given asymmetry of the barriers $\kappa$ the amplitude of the oscillations is not strongly dependent on $V_{0}$. For example, as presented in Fig.~\ref{Fig4}, the amplitudes for $V_{0}=5$ and $V_{0}=10$ are not
much different when the same value of $\kappa=2/5$. In contrast, the frequency of the oscillations is essentially affected by the choice of 
$V_{0}$ and it is larger for stronger $V_{0}$. The latter observation implies that for very large $V_{0}$, experimental detection of oscillations will be very challenging due to the finite resolution of time probes. Simply, to have any realistic chance to detect the effect, a period of the oscillation should not be smaller than the experimental time resolution.

Importantly, it should be pointed here that in our work we do not consider deviations from the exponential decay occurring always for very large times, {\it i.e.} when the decay is characterized by the power-law rather than the exponential one \cite{khalfin,winter,rothe}. In fact, this regime is not well captured in our analysis due to the numerical simplification of the model described in Appendix A. Although going beyond this approximation is straightforward, it highly increases numerical complexity without changing the results in the time ranges we are interested in. Therefore, the discussion on properties of the ratios $\boldsymbol{\Pi}(t)$ and $\boldsymbol{\pi}(t)$ for very long times is beyond the scope of this work.

One can expect that the qualitative features of the results obtained do not significantly depend on the details of the employed decay model. This conviction is justified since the origin of different behavior of $\boldsymbol{\pi}(t)$ and $\boldsymbol{\Pi}(t)$ is ingrained in the fundamental properties of the two-channel decay rather than a particular physical realization. Note that both quantities are described by the same decay width $\beta$ only in the BW limit independently in the underlying model. It means that any deviation from this prediction is a direct manifestation of the non-exponential decay. In other words, as long as the probabilities for the two partial decay channels are not equal, the corresponding functions $\mathrm{P}_{L}(t)$ and $\mathrm{P}_{R}(t)$ approach the respective exponential limits in a slightly different way. Consequently, ratios $\boldsymbol{\Pi}(t)$ and $\boldsymbol{\pi}(t)$ are characterized by slightly different and time-dependent parameters. This is the intuitive reason why the ratios enhance the differences quite independently on the details of a model. This is also one of the reasons why very similar results were obtained in a completely different context in \cite{duecan} in the framework of the Lee model \cite{lee} containing essential simplification when compared to the generalized Winter's model considered here. In contrast to the case studied, in the Lee model it is assumed that there exist the unique {\it unstable} state $|\psi_0\rangle$ decaying to two different subspaces (channels $L$ and $R$) spanned by states $|k,L\rangle$ and $|k,R\rangle$ having the same dispersion relation $\omega(k)$. In such a case the Hamiltonian of the system can be written explicitly in the basis of these states as
\begin{align}
{\cal H}_{\mathtt{Lee}} &=E_0|\psi_{0}\rangle\langle \psi_{0}| +\sum_{\sigma\in\{L,R\}}{\int\limits_{0}^{\infty}} \mathrm{d}k\,\omega(k)|k,\sigma\rangle\langle k,\sigma| \nonumber\\
&+\sum_{\sigma\in\{L,R\}}\int\limits_{0}^{\infty}
\mathrm{d}k\,\Big[f_{\sigma}(k)|k,\sigma\rangle\langle \psi_{0}| + f^*_{\sigma}(k)|\psi_{0}\rangle\langle
k,\sigma|\Big],
\end{align}
where $f_\sigma(k)=\langle k,\sigma|H|\psi_0\rangle$ are transition amplitudes controlling tunneling through the barriers. The non-exponential decay observed in these two, essentially different models, suggests once more that our findings on properties of ratios $\boldsymbol{\pi}(t)$ and $\boldsymbol{\Pi}(t)$ persist model-independent. 

\section{Conclusions}

In this work, we analyzed the general problem of capturing non-exponential
properties in the presence of the two-channel decay process. Taking as a
working horse a very simple dynamical problem of a single particle flowing out
from a leaky box, we examined direct relations between the probabilities of
tunneling to the right and the left as functions of the control parameters. In
this way, we studied relations between partial decays into two distinct
channels in a relatively simple system, which allows for a very accurate
numerical treatment. Since the multiple channel decay of an unstable quantum
state is a very frequent problem in QM and QFT, the results can be important
for our understanding of a broad range of physical phenomena.

The results obtained confirm that in the presence of two decay channels, the
system exhibits a remarkable non-exponential behavior on long time-scales. Even in cases, when the simplest quantities do not reveal any
non-exponential signatures, the inter-channel ratio of probability currents
$\boldsymbol{\pi}(t)$ directly exposes these features. Importantly, this
quantity, although being not the simplest property of the system, is almost
directly measurable in experiments \cite{jochim,fallani,kuzmenko}. Therefore,
it can be viewed as a possible \textit{smoking gun} of
non-exponential decay behavior.

It is worth to point out that the model discussed in this work, although seemingly oversimplified, to some extend can be realized experimentally and give prospects for direct verification of our predictions. State-of-the-art experiments \cite{exp1,exp2,exp3,exp4} with ultra-cold atoms confined in optical traps allow to prepare quasi-one-dimensional uniform box traps where particles are confined. Moreover, outside walls of these traps can be controlled independently and released almost on-demand opening direct routes to realize our model. Another interesting direction of experimental realization is to analyze different nuclei with non-symmetric few-channel decays, for instance, the decay of $\alpha$ particle in large non-spherical nuclei.

From a theoretical point of view, one can easily extend the present work to
more complicated (and more realistic) forms of asymmetric potentials. While
any qualitative differences from the results obtained are not expected, such
studies would help to establish a closer relevance to upcoming experimental
schemes. From the conceptual side, extensions of the results to higher
dimensions are also straightforward. Another promising route for further
explorations is to study analogous systems containing several interacting
particles
\cite{koscikokopinska,fewbody1,fewbody2,fewbody3,fewbody4,fewbody5,fewbody6,fewbody7,fewbody8,fewbody9,sowinski,sowinski2}
and pin down the role of the quantum statistics. Furthermore, the topic should
be also re-investigated in the realm of QFT and shed some fresh light on the
problem of multichannel decays of elementary particles and composite hadrons.

\acknowledgments
FG thanks G.\ Pagliara, M. Piotrowska, and K. Kuli\'{n}ska-Maciejska for
useful discussions. TS acknowledges financial support from the (Polish)
National Science Center under grant No. 2016/22/E/ST2/00555.

\appendix

\section{Numerical approach}

Numerical calculations are performed in the basis of the eigenstates of the
Hamiltonian (\ref{hamiltonadim}) diagonalized numerically on a finite spatial
interval with closed boundary conditions at $x=\pm L$. Everywhere besides the
points $x=\pm1$ the Hamiltonian is equivalent to the Hamiltonian of a free
particle. Therefore, any of its eigenstates can be expressed as following
\begin{equation}
\mathcal{\psi}(x)=\left\{
\begin{array}
[c]{ll}%
A\sin(p(L+x)), & \text{if}\,\,x<-1\\
B\sin(p(L-x)), & \text{if}\,\,x>1\\
C\sin(px)+D\cos(px)\text{,} & \text{if}\,\,|x|\leq1
\end{array}
\right. , \label{inter}%
\end{equation}
where parameters $A$, $B$, $C$, and $D$ are established in such a way that the
wave function fulfills continuity conditions at positions of the left and the
right barrier. These four conditions read
\begin{subequations}
\begin{align}
\lim_{\epsilon\rightarrow0}\left[ \psi(-1+\epsilon)-\psi(-1-\epsilon)\right]
& =0,\\
\lim_{\epsilon\rightarrow0}\left[ \left. \frac{\mathrm{d}}{\mathrm{d}x}%
\psi(x)\right\vert _{-1+\epsilon}-\left. \frac{\mathrm{d}}{\mathrm{d}x}%
\psi(x)\right\vert _{-1-\epsilon}\right]  & =2V_{L}\psi(-1),\\
\lim_{\epsilon\rightarrow0}\left[ \psi(1+\epsilon)-\psi(1-\epsilon)\right]
& =0,\\
\lim_{\epsilon\rightarrow0}\left[ \left. \frac{\mathrm{d}}{\mathrm{d}x}%
\psi(x)\right\vert _{1+\epsilon}-\left. \frac{\mathrm{d}}{\mathrm{d}x}%
\psi(x)\right\vert _{1-\epsilon}\right]  & =2V_{R}\psi(1)
\end{align}
and they lead to the homogenous system of linear equations of the form
$\mathcal{M}\cdot\vec{v}=0$, where $\vec{v}=(A,B,C,D)^{T}$ and
\begin{widetext}
${\cal M}= \left(
\begin{array}{cccc}
\frac{1}{2} p \cos ((L-1) p) & 0 & -\frac{1}{2} p \cos (p)-V_{L} \sin (p) & V_{L} \cos (p)-\frac{1}{2} p \sin (p) \\
0 & -\frac{1}{2} p \cos ((L-1) p) & \frac{1}{2} p \cos (p)+V_{R} \sin (p) & V_{R} \cos (p)-\frac{1}{2} p \sin (p) \\
\sin ((L-1) p) & 0 & \sin (p) & -\cos (p) \\
0 & -\sin ((L-1) p) & -\sin (p) & -\cos (p) \\
\end{array}
\right).$
\end{widetext}In this way the allowed momenta $p_{i}$ and the
corresponding coefficients $\vec{v}_{i}$ are determined. Then, the the
time-dependent wave function is simply given as
\end{subequations}
\begin{equation}
\Psi(x,t)=\sum_{i}\alpha_{i}\exp\left( -itp_{i}^{2}/2\right) \mathcal{\psi
}_{i}(x), \label{expansion}%
\end{equation}
where the expansion coefficients $\alpha_{i}$ are determined by the initial
wave function (\ref{initial}). The accuracy of the final results is easily
controlled (and if needed may be straightforwardly improved) by changing the
number of terms in the expansion (\ref{expansion}). Typically, in our
calculations, we use $3000$ terms and $L=400$-$600$ which is sufficient to
achieve well-converged results avoiding reflections at the walls at $x=\pm L$
for large $t$. The method used assures a full control on accuracy of the final results.


\begin{thebibliography}{99}
\bibitem{footnote2} The problem of adaptation of the general scattering solutions reported in e.g. Ref. \cite{Guilarte} to the problem of decay studied here originates in the lack of orthogonality between left-to-right and right-to-left scattering solution. Finding an orthogonal and complete basis is not an easy task since the orthogonalizing transformation is momentum dependent. More precisely, the problem can be -as expected- solved in the symmetric case: the sum and the difference of the scattering wave solutions are even and odd functions respectively, hence they are orthogonal and the desired basis can be found. However, in the general (and for us crucial) asymmetric case, this is not possible:  for each momentum $k$ a different field transformation should be performed and it is at present unclear how to handle the problem. (Note, the standard Gram-Schmidt method does not apply for states whose norm is infinite). In view of these difficulties, since the numerical solutions that we provide are correct and since the derivation of the analytic solution is, albeit interesting and desirable, not central for our goals, we
decided to leave this task for the future.
\bibitem {footnote}There is a subtle but important difference between the
nondecay and the survival probability at a given time $t$. The former is the
probability that no decay has occurred at $t$, the latter that the particle
occupies the same initial state. In this sense, Eq. (\ref{ndp}) defines the
nondecay probability, while the survival probability is defined as
$P_{\text{surv}}(t)=\left\vert \int_{-1}^{+1}\!\mathrm{d}x\,\Psi_{0}%
(x)\Psi(x,t)\right\vert ^{2}$. Clearly, for the system studied the nondecay
probability is the natural quantity to consider.
\end{thebibliography}
\end{document}